\documentclass[12pt]{article}\usepackage[dvips]{graphicx}
\usepackage{latexsym}
\usepackage{amssymb}

\newcommand{\C}{\mathbb{C}}
\newcommand{\CP}{\mathbb{CP}}
\newcommand{\RP}{\mathbb{RP}}
\newcommand{\R}{\mathbb{R}}

\newcommand{\Z}{\mathbb{Z}}

\renewcommand{\d}{\mathrm{d}}

\def\be{\begin{equation}}
\def\ee{\end{equation}}

\def\Om{\Omega}

\def\om{\omega}

\def\ov{\overline}

\def\p{\partial}

\def\ov{\overline}

\def\sm{\sigma}

\newcommand{\hook}{{\setlength{\unitlength}{11pt}   
                   \begin{picture}(.833,.8)
                   \put(.15,.08){\line(1,0){.35}}
                   \put(.5,.08){\line(0,1){.5}}
                   \end{picture}}}
\def\a{\alpha}
\def\ll{\lambda}

\begin{document}
\pagestyle{plain}

\title{\vskip -70pt
\begin{flushright}
{\normalsize DAMTP-2004-125} \\
\end{flushright}
\vskip 80pt
{\bf Reduced dynamics of Ward solitons} \vskip 20pt}

\author{Maciej Dunajski\thanks{email M.Dunajski@damtp.cam.ac.uk} \\[10pt]
and \\[10pt]
Nicholas S. Manton\thanks{email N.S.Manton@damtp.cam.ac.uk} \\[15pt]
{\sl Department of Applied Mathematics and Theoretical Physics} \\[5pt]
{\sl University of Cambridge} \\[5pt]
{\sl Wilberforce Road, Cambridge CB3 0WA, UK} \\[15pt]}
\date{November 5, 2004} 
\maketitle

\begin{abstract}
The moduli space of static finite energy solutions
to Ward's integrable chiral model is the space  $M_N$ of based rational maps
from $\CP^1$ to itself with degree $N$. 
The Lagrangian of Ward's model gives rise to 
a K\"ahler metric and a magnetic vector potential
on this space. However, the magnetic field strength vanishes, and the 
approximate non--relativistic solutions to Ward's model correspond to a
geodesic motion on $M_N$. These solutions can be compared with exact
solutions which describe non--scattering or scattering solitons.
\end{abstract}

\newpage
\section{Introduction}
\setcounter{equation}{0}
There aren't any known examples of Lorentz invariant equations
admitting soliton solutions in $2+1$ dimensions. The first order 
Yang--Mills--Higgs system proposed by Ward \cite{W88} 
almost does the job: the unknowns $(A, \Phi)$
are a one--form and a function which depend on local coordinates $x^\mu
= (t, x, y)$, and 
take values in ${\bf su}(2)$, the  Lie algebra of $SU(2)$. The metric on
$\R^{2+1}$ is $\d s^2=-\d t^2+\d x^2+\d y^2$. The equations are
\be
\label{Wardeq}
D\Phi=*F.
\ee
Here $F=(1/2)F_{\mu\nu}\d x^\mu\wedge \d x^\nu$ is the 
curvature of a connection $A=A_{\mu}\d x^{\mu}$.
The action of the covariant derivative $D_\mu$ on $\Phi$  is
$D_\mu\Phi=\p_\mu \Phi+[A_\mu, \Phi]$, and
\[
F_{\mu \nu}=[D_\mu, D_\nu]=\p_\mu A_\nu-\p_\nu A_\mu+[A_\mu, A_\nu].
\]
Solutions of (\ref{Wardeq}) are defined 
modulo the gauge transformations
\[ 
A_\mu\longrightarrow\hat{A}_\mu=gA_\mu g^{-1}-(\p_\mu g)g^{-1},\qquad
F_{\mu\nu}\longrightarrow\hat{F}_{\mu\nu}=gF_{\mu\nu}g^{-1}, 
\qquad \Phi\longrightarrow\hat{\Phi}=g\Phi g^{-1},
\]
where $g(x^{\mu})\in SU(2)$. 

This system is integrable in more than one way: it arises as a
reduction of $2+2$ dimensional self--dual Yang--Mills equations by a non--null
translation, it possesses an infinite number of conserved currents
and so forth. 
It can not however be regarded as a genuine soliton system, because
the energy functional associated to the Lagrangian
\[
\int \Big\{\frac{1}{2}\mbox{Tr}(F_{\mu\nu}F^{\mu\nu})-
\mbox{Tr}(D_{\mu}\Phi D^{\mu}\Phi)\Big\}\d x\d y
\]
is not positive definite and its density vanishes on 
all solutions to (\ref{Wardeq}) (the given Lagrangian differs from the 
standard Yang--Mills--Higgs Lagrangian by the 
relative sign between the two terms; the second order 
Euler--Lagrange equations are satisfied by solutions to the first
order system (\ref{Wardeq})).

There exists a positive functional associated to (\ref{Wardeq}). To
see it, note that the equations (\ref{Wardeq}) arise as the
integrability conditions for an overdetermined system of linear
equations
\be
\label{laxpair}
(D_y+D_t-\ll(D_x+\Phi))\psi=0,\qquad
(D_x-\Phi-\ll(D_t-D_y))\psi=0,
\ee
where $\psi$ is an $SU(2)$--valued function of $(t, x, y)$ and a complex
parameter $\ll$. 
The integrability conditions for (\ref{laxpair}) 
imply the existence of a gauge $A_t=A_y$ and  $A_x=-\Phi$, and a matrix
$J:\R^{2+1}\longrightarrow SU(2)$ such that 
\[
A_t=A_y=\frac{1}{2}J^{-1}(J_t+J_y), \qquad
A_x=-\Phi=\frac{1}{2}J^{-1}J_x,
\]
and equations (\ref{Wardeq}) become
\be
\label{chiral}
(\eta^{\mu\nu}+V_\a\varepsilon^{\a\mu\nu})\p_{\mu}(J^{-1}\p_{\nu}J)=0.
\ee
Here $\eta^{\mu\nu}=$diag$(-1, 1, 1)$ is the inverse of the Minkowski
metric, $\varepsilon^{\a\mu\nu}$ is the  alternating tensor with 
$\varepsilon^{012}=1$, and $V_\a=(0, 1, 0)$.
The energy functional associated with (\ref{chiral}) is
\be
\label{energy}
E=\int\frac{1}{2}\delta^{\mu\nu}\mbox{Tr}(\p_\mu J\p_\nu
J^{-1})\d x\d y=\int{\cal E}\d x\d y,
\ee
and it is conserved. 
This functional is positive definite, but it came at the price of
losing the Lorentz invariance: any choice of a constant space--like
vector $V_\a$ fixes a direction in $\R^{2+1}$. Finiteness of $E$ 
is ensured  by imposing the boundary condition (valid for all $t$)
\be
\label{assympt}
J=J_0+J_1(\theta)r^{-1}+O(r^{-2})\qquad 
\mbox{as}\qquad r\longrightarrow \infty,\qquad x+iy=re^{i\theta}.
\ee

The integrability of equations (\ref{Wardeq}), or equivalently of 
(\ref{chiral}), allows a  construction of explicit static and also 
time--dependent 
solutions by twistor or inverse--scattering methods \cite{W88,W90}.
Static solutions are described by rational maps and may be identified with
lumps or sigma model solitons. There are time--dependent solutions 
with non--scattering solitons \cite{W88}, and also solitons that 
scatter \cite{W95}.

In this paper we choose a different route and seek slow--moving
solitons using a modification of the  geodesic approximation \cite{M82} 
which may involve 
a background magnetic field in the moduli space of static solutions. 
The  argument is based on the analogy with a particle in $\R^n$
moving in a potential $U$ and coupled to a magnetic vector potential 
${\bf A}({\bf q})$; the Lagrangian is
\[
L=\frac{1}{2}|{\bf \dot{q}}|^2+{\bf A}\cdot {\bf \dot{q}}-U({\bf q}),
\]
where $U:\R^n\rightarrow \R$ is a potential whose minimum value is $0$. 
The equilibrium positions are on a subspace $M\subset \R^n$
given by $U=0$. If the kinetic energy of the particle is small, and
the
initial velocity is tangent to $M$, the  exact motion will be
approximated by a motion on $M$ with the Lagrangian $L'$ given by a
restriction of $L$ to $M$ 
\be
\label{reduced_lag}
L'=\frac{1}{2}h_{jk}\dot{\gamma}_j\dot{\gamma}_k+A_j\dot{\gamma}_j.
\ee
Here, the $\gamma$s are local coordinates on $M$,
and  the metric $h$ and the one--form $A_j\d \gamma_j$ are induced on
$M$ from the Euclidean inner product and the magnetic vector potential 
${\bf A}$  respectively. 
If for example $U=(1-r^2)^2$, where $r=|{\bf q}|$, 
and the magnetic field is constant,
then  the  motion with small energy is approximated by the motion
on the unit sphere in $\R^n$ where trajectories are small circles,
that is, a circular motion at $r=1$ with constant speed. 

In the absence of the magnetic term we
expect the true motion to have small oscillations in the direction 
transverse to $M$, with the approximation becoming exact at the limit
of zero initial velocity. The presence of a magnetic force may in some
cases balance the contribution from a centrifugal force so that the
oscillations do not occur, and the 
exact motion is confined to $M$. 

The dynamics of finite energy solutions to (\ref{chiral}) will be put
in this framework with $\R^n$ replaced by an infinite--dimensional
configuration space of the field $J$, and $M$ replaced by $M_N$ 
(the moduli space of rational maps from $\CP^1$ to itself with degree $N$),
the important point being that the static solutions to (\ref{chiral})
give the absolute minimum of the potential energy for given $N$.
The time--dependent solutions to (\ref{chiral}) with
small total energy (hence small potential energy) above the absolute minimum 
will be approximated by a sequence of static states, 
that is a motion in $M_N$. 

This comes down to three steps:
\begin{enumerate}
\item Construct finite--dimensional families of static solutions to 
(\ref{chiral}) with
finite energy.
\item Allow time--dependence of the parameters, and read off the metric $h$
and the magnetic one--form $A$  
on the moduli space from the Lagrangian for $J$. Investigate whether $A$ has a
non--vanishing or vanishing magnetic two--form $F=\d A$.
Some of the parameters may have to be fixed artificially to ensure
that this metric is complete, and all tangent vectors have finite length.
\item The geodesic motion, possibly with magnetic forcing, should then 
approximate the slow (non--relativistic) motion of rational map, or 
lump solutions to (\ref{chiral}).
\end{enumerate}

In the next section we shall find  the metric and the one--form
to be 
\be
\label{metric_one_form}
h_{jk}=8\int_{\R^2}\frac{|\p_jf\p_kf|}{(1+|f|^2)^2}\d x\d y,
\qquad 
A_j=4\pi\int_{\R^2}\frac{ \mbox{Re}(\p_zf\p_j\ov{f})}{(1+|f|^2)^2}\d x\d y.
\ee
Here $f=f(z)$ is a rational meromorphic function of $z$, which depends
on real parameters $\gamma_j$, and $\p_j f=\p f/\p\gamma_j$. We
shall show that the magnetic two--form $F=\d A$ in fact vanishes on the
moduli space\footnote{It is worth remarking that
even magnetically forced geodesic motion can be regarded 
as a geodesic motion on an $S^1$--bundle ${\cal L}\rightarrow M_N$
equipped with a connection $\om$ with curvature $F$. 
In a local trivialisation 
$\om=\d\theta+A$, where $\theta\in S^1$ is a coordinate along the fibres.
The motion in a magnetic field on $M_N$ given by 
(\ref{reduced_lag}) is geodesic on ${\cal L}$ with respect to 
a Kaluza--Klein metric $\hat{h}=h+\om\otimes\om  $. This 
can be verified by writing the Euler--Lagrange equations of
\[
\hat{L}=\frac{1}{2}h(\dot{\gamma}, \dot{\gamma})+\frac{1}{2}(\dot{\theta}+
h(A,\dot{\gamma}))^2.
\]
One equation implies that $\dot{\theta}+h(A, \dot{\gamma})=C$ is a constant
which can be chosen so that 
the remaining equations coincide with those obtained 
from (\ref{reduced_lag}).}. 
\section{Reduced dynamics}
\setcounter{equation}{0}
All static solutions to (\ref{chiral}) are just chiral fields on
$\R^2$, i.e. solutions to
\be
\label{static}
\ov{\p_z}(J^{-1}\p_z J)+\p_z(J^{-1}\ov{\p_z}J)=0
\ee
where $z=x+iy$, and $\p_z=\p/\p z$.

It is convenient at this point to fix $J_0$ to lie in the equator 
$S^2\subset SU(2)$ given by $J_0^2=-{\bf 1}$. We shall choose 
$J_0 = i\sigma^1$, where $\sigma^1, \sigma^2, \sigma^3$ are the Pauli matrices.
It has been shown in \cite{U89} that all chiral fields with finite energy
and satisfying our boundary conditions can be globally conjugated into 
the so--called 1--uniton solutions
\be
\label{uniton}
J=
\frac{i}{1+|f|^2}
\left (
\begin{array}{cc}
1-|f|^2&2f\\
2\ov{f}&|f|^2-1
\end{array}
\right ),
\ee
where the holomorphic function $f$ is rational in $z$ and 
$f(z)\longrightarrow 1$ as $|z|\longrightarrow \infty$. General
solutions to the chiral equations are $SU(2)=S^3$ valued, but the 
1--uniton (\ref{uniton}) everywhere takes values in the equator 
$J^2=-{\bf 1}$, 
and is in effect a based harmonic map from a two--sphere (conformal
compactification of $\R^2$) to itself. All such maps are classified by
integer winding numbers $N$ with values in $\pi_2(S^2)=\Z$. This integer
is precisely the degree of $f$: for a given
$N$, $f$ is of the form
\be
\label{rational}
f(z)=\frac{p(z)}{q(z)}=
\frac{(z-q^1)...(z-q^N)}{(z-q^{N+1})...(z-q^{2N})},
\ee
and the map $f$ is an $N$--fold covering of the target $\CP^1 = S^2$.

Let $M_N\subset \C^{2N}$ be the moduli space of 1--unitons 
of degree $N$. This space
consists of all based rational functions of degree $N$ 
(we assume that the denominator and the numerator in (\ref{rational}) 
have no common factors) and has  real dimension $4N$. 
Let the parameters of the solution be denoted collectively by 
$\gamma=(q, \ov{q})$, and
let $J(\gamma)$ be the corresponding solution to (\ref{static}) 
(this solution also depends on $x,y$). Let $\gamma(t)$ be a path in $M_N$.
Then $J(\gamma(t))$ is not in general a solution to the time--dependent
eq.(\ref{chiral}), but for $t=0$ it provides   initial data for 
$J$ and its derivative. The initial velocity
\[
\dot J|_{t=0}=\frac{\p J(\gamma(t))}{\p \gamma_i}\dot{\gamma}_i\biggr|_{t=0}
\] 
is tangent to $M_N$, and is a linearised solution to (\ref{static}). 
If this initial velocity is small, the true
dynamical motion will remain close to $M_N$ because of the conservation of
energy. In the limiting case, when the velocity remains small, the motion
will be governed by a Lagrangian of the form (\ref{reduced_lag}).

To find this moduli space Lagrangian we need the
action for $J$, which is a sum of a standard part quadratic
in the derivatives of $J$, and a Wess--Zumino--Witten (WZW) 
like term \cite{WZ71, Wi84}. This involves an extended field 
\[
\hat{J}:\R^{2+1}\times[0,1]\longrightarrow SU(2)
\]
such that $\hat{J}(x^{\mu}, 0)$ is a constant group element, which we
take to be the identity element, and
$\hat{J}(x^{\mu}, 1)=J(x^{\mu})$. That is to say $\hat{J}$ is defined in the
interior of a cylinder which has the space--time as one of its
boundary components. The action is 
\be
\label{action}
S=S_C+S_M,
\ee
where
\begin{eqnarray*}
S_C&=&-\int_{\R^2}\int_{t_1}^{t_2}\frac{1}{2}
\mbox{Tr}((J^{-1}J_t)^2-(J^{-1}J_x)^2-(J^{-1}J_y)^2)\d t\d x\d y,\\ 
S_M&=&-\int_{\R^2}\int_{t_1}^{t_2}\int^1_0 
\frac{1}{3}V_p\varepsilon^{pqrs}
\mbox{Tr}(\hat{J}^{-1}\p_q \hat{J} 
\hat{J}^{-1}\p_r \hat{J} 
\hat{J}^{-1}\p_s \hat{J})\d \rho\d t\d x\d y.
\end{eqnarray*}
The indices $p, q, r, s$ take values
$0,1,2,3$, where $x^3=\rho$ and $V=(0, 1, 0, 0)$. 
In \cite{IZ} it was demonstrated that the variation of the action
does not depend on the choice of the extension $\hat{J}$, and leads to
the Ward model equation (\ref{chiral}).

The kinetic part of the effective Lagrangian (\ref{reduced_lag}) 
can be found as follows: Given a 
path $\gamma(t)$ in $M_N$ we define the kinetic
energy at time $t$ by
\[
T[J]=-\frac{1}{2}\int\mbox{Tr}(J^{-1}J_t)^2\d x\d y.
\]
Substituting for $J$ from (\ref{uniton}) yields
\be
\label{Mmetric}
h(\dot{\gamma}(t), \dot{\gamma}(t))= 2T[J]
=\int\frac{8|\dot{f}|^2}{{(1+|f|^2)}^2}\d x\d y,
\ee
where $f$ depends on $\gamma(t)$, and hence $\dot{f}$ depends on $\gamma$
and $\dot{\gamma}$. Expressing $\dot{f}$ as $\p_j f \, \dot{\gamma}_j$
verifies the first part of (\ref{metric_one_form}).

Notice that there is no potential term in the Lagrangian (\ref{reduced_lag}), 
since for the static solutions (\ref{uniton}), 
the potential energy part of $S_C$ is minimised by the degree of the 
rational function $f$:
\be
\label{uniton_en}
{E}=-\frac{1}{2}\int\mbox{Tr}((J^{-1}J_x)^2+(J^{-1}J_y)^2)\d x\d y=
 4i\int_{S^2}\frac{\d f\wedge \d \ov{f}}{(1+|f|^2)^2} =8N\pi,
\ee
where in the last integral we have used the fact that the 
solution (\ref{uniton})
extends to a one--point compactification of $\R^2$, and that the rational 
function (\ref{rational}) is an $N$--fold covering of the
two--sphere. This constant potential energy can be dropped.

The analysis of slowly moving lumps in the $\CP^1$ model leads to
an identical expression for the kinetic energy \cite{W85}. 
(This was to be expected, because the conserved functional (\ref{energy})
is identical to that of the non--integrable chiral equation obtained by
setting $V_\a=0$.) The slow dynamics could however be different for these
two models: trajectories of slow moving $\CP^1$ lumps are just the geodesics
of $h$, but the motion of lumps of (\ref{chiral}) would be affected 
by any magnetic field $F$ on the moduli space.

Using the cyclic property of the trace we can simplify 
the integrand of the WZW term $S_M$ to
\[
{\cal L}_M=\mbox{Tr}([\hat{J}^{-1}\hat{J}_y,\hat{J}^{-1} \hat{J}_t]
\hat{J}^{-1}\hat{J}_\rho).
\]
One can now understand the vanishing of $F$. The
variation of $S_M$ involves the integral over 
$\R^{2}\times[t_1,t_2]$ of 
$\mbox{Tr}(J^{-1}\delta J [J^{-1}J_t,J^{-1}J_y])$. If $J$ is
restricted to the equator $J^2=-{\bf 1}$, for all $x,y$ and $t$, then
$J^{-1}\delta J$, $J^{-1}J_t$ and $J^{-1}J_y$ lie in a two--dimensional 
subspace of ${\bf su}(2)$ at any given space--time point, 
and the trace above vanishes. In the moduli
space approximation, $J$ is restricted in this way, and the variation
of the action under a change of path in the moduli space (with fixed
endpoints) has no contribution from the WZW term. There is therefore
no magnetic force in the reduced equation of motion.

Despite this, we shall calculate the `magnetic' one--form $A$ in 
(\ref{reduced_lag}) from the WZW term. We make the ansatz 
\[
\hat{J}(x^{\mu}, \rho)=\cos{g(\rho)}{\bf 1}+\sin{g(\rho)}J,
\]
where $J$ is the static solution given by (\ref{uniton}),
and $g(\rho)$ is any smooth function on the interval $[0, 1]$ such that
$g(0)=0, g(1)=\pi/2$. 
This, with the help of $J^2=-{\bf 1}, J^*=-J$ and 
some algebra, reduces ${\cal L}_M$ to 
\[
{\cal L}_M=\sin^2{g(\rho)}\frac{\d g({\rho})}{\d \rho}
\mbox{Tr}{(J[J_y, J_t])},
\]
and the magnetic one--form on the moduli space can be read off from 
this Lagrangian density 
\[
\int^{t_2}_{t_1}A_i(\gamma){\dot{\gamma}}^i\d t=
\int_{t_1}^{t_2}\int_{\R^2\times[0,1]}{\cal L}_M \d x\d y\d \rho\d t.
\]
Using
\[
\mbox{Tr}\left(J\;\left[\frac{\p J}{\p f}, \frac{\p J}{\p\ov{f}}\right]\right)
=-\frac{8i}{{(1+|f|^2)}^2}
\]
we find
\be
\label{magA}
A_i(\gamma)\dot{\gamma}^i=\frac{\pi}{4}\int \mbox{Tr}(J[J_y, J_t])\d x\d y
=4\pi\int\frac{\mbox{Re}{(f_z\dot{\ov{f}})}}{(1+|f|^2)^{2}}\d x\d y,
\ee
which establishes (\ref{metric_one_form}). 

\subsection{K\"ahler and magnetic potentials}
\label{kmpot}
The metric (\ref{Mmetric}) is known to be K\"ahler \cite{R88,S03}, with 
the holomorphic coordinates defined to be any  functions of
$(q^{\a})=(q^1, ..., q^{2N})$.
Let $\d=\p+\overline{\p}$ be a decomposition of
the total derivative on this K\"ahler manifold, and let $f=p(z)/q(z)$ be 
given by (\ref{rational}). Then
\[
A=\p \Big(\int\chi\d x\d y\Big)+
\overline{\p}\Big(\int\ov{\chi}\d x\d y\Big),\qquad\mbox{where}\qquad
\chi =2\pi\frac{\p}{\p \ov{z}}\Big(\ln{(1+|f|^2)}\Big),
\]
so both the magnetic field $F=\d A$, and the metric $h$, can be
written in terms of scalar potentials
\[
F=i\p\wedge\ov{\p}  \Om_F \in \Lambda^{(1, 1)}(M_N),\qquad h=
\frac{\p^2\Om_h}{\p q^{\a}\p{q}^{\ov\beta}}\ dq^\a\d q^{\ov{\beta}},
\]
where
\be
\label{magF}
\Om_F=-2\pi \int_{\R^2}\frac{\p}{\p y}\ln{(|p|^2+|q|^2)}\d x\d y,
\qquad \Om_h=8\int_{\R^2}\ln{(|p|^2+|q|^2)}\d x\d y,
\ee
and we have used the freedom of adding any holomorphic or antiholomorphic
functions of $q^\a$ to the potentials  $\Om_F$ and $\Om_h$.

The equations of motion in the moduli space approximation in 
holomorphic coordinates are
\[
\ddot{q}^{\a}+\Gamma^{\a}_{\beta\gamma}\dot{q}^{\beta}\dot{q}^{\gamma}
=-h^{\a\ov{\beta}}F_{\gamma{\ov\beta}}\dot{q}^{\gamma}
\]
(and the complex conjugate of this), where
\[
F_{\a\ov{\beta}}=2i\frac{\p^2\Om_F}{\p q^{\a}\p q^{\ov{\beta}}}, \qquad
h_{\a\ov{\beta}}=\frac{\p^2\Om_h}{\p q^{\a}\p q^{\ov{\beta}}}, \qquad
\Gamma^{\a}_{\beta\gamma}=h^{\a\ov{\delta}}
\frac{\p h_{\beta\ov{\delta}}}{\p q^{\gamma}},
\]
and formulae (\ref{magF}) imply 
that $f(z)$ and $f(z)^{-1}$ give rise to the same dynamics
on moduli space.

However we shall now show  that a suitable regularisation of the 
 magnetic scalar potential leads to the vanishing of 
$F$. 
Set
\be
\label{bound_pq}
p=z^n+az^{n-1}+... \,,\qquad q=z^n+bz^{n-1}+...\,,
\ee
and consider $\Om_F$ in (\ref{magF}) with the integral over $\R^2$ replaced
by the integral over an annulus 
$D(\epsilon, R)=\{z=r\exp{(i\theta}), \epsilon<r<R\}$ 
bounded by circles $C_R$ and $C_\epsilon$ of radii $R$ and $\epsilon$ respectively.
We will regularize the integrand by subtracting $\p/\p
y(\ln{2|z|^{2n}})$, as this term does not contribute to $F$.
The application of Green's theorem gives
\begin{eqnarray*}
\Om_F&=&\lim_{R\rightarrow \infty}\lim_{\epsilon\rightarrow 0}
\int_{D(\epsilon, R)}2\pi \frac{\p}{\p
  y}\ln{\Big(\frac{|p|^2+|q|^2}{2|z|^{2n}}\Big)}\d x\d y\\
&=&\lim_{R\rightarrow \infty}\lim_{\epsilon\rightarrow 0}
\Big(\oint_{C_R}-\oint_{C_{\epsilon}}\Big)2\pi
\ln{\Big(\frac{|p|^2+|q|^2}{2|z|^{2n}}\Big)}\d x\\
&=&\lim_{R\rightarrow \infty}\int_0^{2\pi}2\pi
R\ln{\left(1+\frac{c_1}{R}+\frac{c_2}{R^2}+...\right)}\sin{\theta}\d\theta\\
&&-\lim_{\epsilon\rightarrow
  0}\int_0^{2\pi}2\pi\epsilon(\ln{(|p|^2+|q|^2)}-2n\ln{\epsilon}-\ln{2})
\sin{\theta}
\d \theta
\end{eqnarray*}
where $c_a$ depend on $\theta$ and 
\[
c_1=\frac{1}{2}(a+b)\exp{(-i\theta)}
+\frac{1}{2}(\ov{a}+\ov{b})\exp{i\theta}.
\] 
The second limit vanishes and the
first term can be expanded for large $R$ to give
\be
\label{OmF0}
\Om_F=\lim_{R\rightarrow
  \infty}
\Big(2\pi\int_0^{2\pi}c_1\sin{\theta}\d\theta
+O\Big(\frac{1}{R}\Big)\Big)=2\pi^2\;\mbox{Im}\;(a+b) ,
\ee
which is a sum of  holomorphic and antiholomorphic functions on $M_N$.
Therefore 
\[
F=0.
\]

We can give a deeper geometrical interpretation of the metric and magnetic 
field on $M_N$.
The static solutions take values in the K\"ahler manifold 
$S^2\subset S^3$ and (following the argument of Ruback \cite{R88})
the K\"ahler structure on $M_N$ is induced from $\CP^1$ as follows.
Let 
\[
p=(x,y), \qquad  X\in T_{f}S^2,\qquad
X(p)=\frac{\d}{\d t}f(\gamma(t), p)|_{t=0}.
\]
If $(\hat{h}_{f(p)}, \hat{\cal J}_{f(p)})$
is the 
standard K\"ahler structure  on $T_{f(p)}S^2$, then 
\[
h_f(X, X)=\int_{\R^2}\hat{h}_{f(p)}(X(p), 
X(p))\d x\d y, \qquad
({\cal J}_f(X))(p)=\hat{\cal J}_f(p)(X(p))
\]
give a metric and an almost complex structure on $T_{\gamma} M_N$. 
It can be formally shown
that $(h, \cal J)$ is in fact a K\"ahler structure.

The magnetic one--form $A_f\in {T^*}_{\gamma}M_N$ 
given by (\ref{metric_one_form}) can be similarly
constructed in terms of a one--form $\hat{A}$ which is
dual to the push--forward of the vector 
field $V$  (compare (\ref{chiral})) from $\R^{2+1}$. Explicitly
\[
f_*(V)=f_*(\p/\p x)=\frac{\p f}{\p z}\frac{\p}{\p f}
+\frac{\p \ov{f}}{\p \ov{z}}\frac{\p}{\p \ov{f}}, \qquad
\hat{A}=\hat{h}(f_*(V), ...)=
8\frac{f_z\d \ov{f}+\ov{f_z}\d f}{(1+|f|^2)^2}
\]
and
\[
X\hook A_f=\int_{\R^2}
(X(p)\hook \hat{A}_{f(p)})\d x \d y  = 
8\int_{\R^2} \dot{\gamma}_j
\frac{f_z\ov{\p_j{f}}+\ov{f_z}\p_j f}{{(1+|f|^2)}^2} \d x \d y,
\]
which coincides with (\ref{metric_one_form}) up to a constant factor if 
$X=\dot{\gamma}_j\p/\p \gamma_j$.

To understand the appearance of the one--form $A$ in this context, one
may also consider a connection on $SU(2)$ (the full target space 
of $J$)
which holds the left--invariant vectors covariantly constant. 
This connection is flat, but necessarily has
torsion which is totally antisymmetric and therefore gives rise to a 
closed three--form $T$. Let $B$ be a locally defined 
two--form such that $T=\d B$. The magnetic  term $S_M$ in the action 
(\ref{action}) is proportional to the integral of $f^*B$ over 
the space-time. Its variation vanishes if $B$ is restricted to
the equatorial two--sphere in $SU(2)$.

\section{Reduced dynamics of the $K$--equation.}
\setcounter{equation}{0}
In this section we shall carry out the moduli space approximation in 
a different potential formulation of (\ref{Wardeq}). 

Choose the familiar gauge 
$A_y=A_t, A_x=-\Phi$. The vanishing of the term proportional to $\ll$ in
the compatibility conditions (\ref{laxpair}) implies the existence of
$K:\R^{2+1}\longrightarrow {\bf su}(2)$ such that
\[
A_y=A_t=\frac{1}{2}K_x, \qquad A_x=-\Phi=\frac{1}{2}(K_t-K_y).
\]
The $0$th order term in the compatibility conditions now yields
\be
\label{Keq}
K_{tt}-K_{xx}-K_{yy}+[K_x, K_t-K_y]=0.
\ee
The relation between $K\in{\bf su}(2)$ and $J\in SU(2)$ is 
\[
K_x=J^{-1}(J_t+J_y), \qquad  K_t-K_y=J^{-1}J_x,
\]
and exhibits a duality between the two formulations: the compatibility 
condition $K_{xt}-K_{xy}=K_{tx}-K_{yx}$ yields the field equation 
(\ref{chiral}).

The $K$--equation (\ref{Keq}) admits a Lagrangian formulation with the
Lagrangian density
\be
\label{KLagr}
-\mbox{Tr}\Big(\frac{1}{2}((K_t)^2-(K_x)^2-(K_y)^2)
-\frac{1}{3} K[K_x, K_t-K_y]\Big).
\ee
For the time--independent solutions we have
$
J^{-1}J_z=-iK_z,\; J^{-1}J_{\ov{z}}=iK_{\ov{z}},
$
and  integrating these equations gives, surprisingly, 
\[
K=J
\]
where $J$ is given by (\ref{uniton}).
This makes sense because
the unit two--sphere in the Lie
algebra ${\bf su}(2)$ may be identified with the equatorial two--sphere in 
$SU(2)$.  
In general,
$
J=a_0{\bf 1}+i{\bf a}\cdot {\bf \sigma},
$
where $a_0$ and  ${\bf a}\in \R^3$ depend on $(t, x, y)$. For the 
static solution (\ref{uniton}), $a_0=0$ and $|{\bf a}|=1$ so
$J^2=-{\bf 1}, \mbox{Tr}(J)=0$,  
which is the equatorial condition. But this means
that $K=J$ lies in the  Lie algebra, and moreover 
$K\in S^2\subset\R^3\cong{\bf su}(2)$.

The energy densities of static
solutions to (\ref{chiral}) and  (\ref{Keq}) are proportional but not equal 
as
\[
3\;\mbox{Tr}\Big(\frac{1}{2}(K_x^2+K_y^2)-\frac{1}{3}K[K_x, K_y]\Big)=
\mbox{Tr}\Big(\frac{1}{2}(J^{-1}J_x)^2+\frac{1}{2}(J^{-1}J_y)^2\Big).
\]
The potential energy in the $K$--formulation is therefore $8N\pi/3$
and again can be dropped.

Now consider the slow--motion approximation, where $K$ is expressed in 
terms of a rational holomorphic function $f$
which depends on time through the collection of $4N$ parameters 
$\gamma(t)$. Then the kinetic energy 
term $-(1/2)\int\mbox{Tr}(K_t)^2\d x\d y$  gives rise to  the metric 
(\ref{Mmetric}) (because $K_t^2=(J^{-1}J_t)^2$).
We conclude that the K\"ahler structures on the moduli spaces of static 
solutions to (\ref{chiral}) and (\ref{Keq}) are the same.

The term in (\ref{KLagr}) of first order in $K_t$ implies that there 
is a magnetic one--form on the space of fields $K$. This is given by 
\begin{eqnarray}
\label{magAK}
{(A_K)}_j(\gamma)\dot{\gamma}^j&=&\frac{1}{3}\int \mbox{Tr}(K[K_x, K_t])\d x\d y
=\frac{16}{3}\int\frac{\mbox{Im}{(f_z\dot{\ov{f}})}}{(1+|f|^2)^{2}}\d x\d y\\
&=&i\p \Big(\int\chi\d x\d y\Big)-i\overline{\p}\Big(\int\ov{\chi}\d x\d y\Big),\quad\mbox{where}\quad
\chi =\frac{8}{3}\frac{\p}{\p \ov{z}}
\Big(\ln{(1+|f|^2)}\Big)\nonumber,
\end{eqnarray}
and $F=i\p\wedge\ov{\p} \Om_{F_K}$
where
\[
\Om_{F_K}=-\frac{8}{3} \int_{\R^2}\frac{\p}{\p x}\ln{(|p|^2+|q|^2)}\d x\d y.
\]
The scalar potentials for the magnetic two--form 
are therefore different in the $J$ and $K$ formulations. 
Nevertheless the limiting procedure described in the last section also applies
in this case (with $y$ replaced by $x$) leading to 
\[
\Om_{F_K}=-\frac{8}{3}\pi\mbox{Re}(a+b).
\]
As before, the magnetic two--form
vanishes when restricted to the moduli space of finite energy static solutions.
\section{Examples}
\setcounter{equation}{0}
In the moduli space approximation $J$ stays on the equatorial 
$S^2\subset S^3$, and 
the lumps are located where $J$ departs from its asymptotic value
(\ref{assympt}). In these regions the energy density of (\ref{energy}) 
attains its local maxima. The velocities of the lumps are the velocities of
these local maxima.

The charge one solution is given by
\be
\label{charge1}
f=\a+\frac{\beta}{z+\gamma},
\ee
and we need to fix $\a$ and $\beta$ in order for (\ref{Mmetric})
to be well defined. Choosing  $\a=0, \beta=1, \gamma=\gamma(t)$,
and setting $\gamma(t)=R(t)\exp{(i\theta(t))}$
we find the metric and the one--form
\[
h=8\pi(\d R^2+R^2\d \theta^2), \qquad
\qquad A={4\pi^2}\d(R\cos{\theta}).
\]
Therefore the metric is flat, and the motion is along straight lines,
$\gamma(t) = -vt$,
because $\d A=0$ does not contribute to the Euler--Lagrange equations.
The energy density is approximated by
\[
{\cal E}=(1+|z-vt|^2)^{-2}.
\]

Next we look at the charge two case\footnote{We remark that the boundary 
condition for $f$ in these examples is $f(z)\longrightarrow 0$ 
as $|z|\longrightarrow \infty$, and hence $J_0 = i\sigma^3$.  
The conclusion $F=0$ does not depend on these boundary conditions.}
\be
\label{charge2}
f=\a+\frac{\beta z+\gamma}{z^2+\delta z+\kappa}.
\ee
The corresponding metric was constructed by Ward \cite{W85}. The parameters
$\a, \beta$ have to be fixed to ensure finiteness of kinetic energy,
and $\delta$ can be set to $0$  by exploiting the translational 
invariance of (\ref{uniton}). Moreover
the M\"obius transformations can be used to ensure  $\a=0, \beta\in\R$, and
here Ward makes an additional choice $\beta=0$. The resulting metric
is therefore defined on four--dimensional leaves of a foliation of $M_2$, 
with local coordinates $(\gamma, \ov{\gamma}, \kappa, \ov{\kappa})$.
The K\"ahler potential is given by
\[
\Om_h=-4\pi|\kappa|+\pi|\gamma|\int_0^{\pi/2}
\sqrt{1+|\kappa/\gamma|^2\sin^2{\theta}}\;\d \theta. 
\]
The structure is invariant under  the torus action and a homothety 
\[
\gamma\rightarrow \exp{(i\tau_1)}\gamma,\qquad
\kappa\rightarrow \exp{(i\tau_2)}\kappa, \qquad
|\gamma|^2+|\kappa|^2\rightarrow \tau_3(|\gamma|^2+|\kappa|^2).
\]
\section{Comparision with exact solutions}
\setcounter{equation}{0}
One method  \cite{W88} of constructing explicit solutions is based on the
associated linear problem (\ref{laxpair}). 
Let $\psi(x^\mu, \ll)$ be 
the fundamental solution to the Lax pair (\ref{laxpair}) (think of $\psi$
as a $2\times 2$ matrix), and let $u=(t+y)/2, v=(t-y)/2$. 
Then 
\begin{eqnarray}
\label{RHzero}
A_u-\ll(A_x+\Phi)&=&[-\p_u \psi+\ll\p_x \psi]\psi^{-1}\nonumber\\
A_x-\Phi-\ll A_v&=&[-\p_x\psi+\ll\p_v \psi] \psi^{-1},
\end{eqnarray}
and in the gauge leading to (\ref{chiral}), $J(x, y, t)=\psi(x, y, t, 0)^{-1}$.
This can be an effective method of finding solutions (also known as the `Riemann problem with zeros'), 
if we know $\psi(x^{\mu}, \ll)$ in the first place.
One class of solutions can be obtained by 
assuming that 
\[
\psi={\bf 1}+\sum_{k=1}^n\frac{M_k(x, y, t)}{\ll-\mu_k},\qquad \mu_{k}=\mbox{const}.
\]
The unitarity condition 
$\psi(x^{\mu}, \ll)\psi(x^{\mu}, \ov{\ll})^*= {\bf 1}$ implies 
$\mbox{rk}\;M_k=1$, and demanding that the RHS of (\ref{RHzero}) is
linear in $\ll$ (like the LHS) yields $M_k=M_k(\om_k)$, where
\[
\om_k=u\mu_k^2+x\mu_k+v.
\]
Finally (see \cite{W88} for details)
\be
\label{ward_sol}
(J^{-1})_{\a\beta}=
\chi^{-1/2}(\delta_{\a\beta}+\sum_{k,l}\mu_k^{-1}(\Gamma^{-1})^{kl}
{\ov{m}_\a}^l{m_\beta^k}).
\ee
Here
\[
\Gamma^{kl}=\sum_{\a=1}^2(\ov{\mu}_k-\mu_l)^{-1}{\ov{m}_\a}^k{{m_\a}^l},
\qquad
\chi=\prod_{k=1}^n\frac{\ov{\mu}_k}{\mu_k},
\]
and ${m_\a}^k=(1, f_k)$.

The soliton solutions correspond to rational functions  $f_k(\om_k)$.
To recover the static solution (\ref{uniton}) put $n=1, \mu=i$. 
The static $N$ lumps are positioned at $(q^1, ..., q^N)$, as the maxima
of ${\cal E}$ occur at these points.
For $\mu\neq\pm i$ there is time dependence, and $n>1$ corresponds to $n$ solitons moving with different velocities which however do not scatter. 

The solution (\ref{ward_sol}) with $n=1$ and $\mu_1=m\exp{i\theta}$ is given by
\be
\label{one_soliton}
J_1=\frac{1}{1+|f|^2}
\left (
\begin{array}{cc}
e^{i\theta}+e^{-i\theta}|f|^2&2i\sin{\theta}\;f\\
2i\sin{\theta}\;\ov{f}&e^{-i\theta}+e^{i\theta}|f|^2
\end{array}
\right ),
\ee
where $f=f(u\mu^2+x\mu+v)$ is a holomorphic, rational function.
The energy density 
\[
{\cal E}=2\sin^2{\theta}\frac{(1+m^2)^2|f'|^2}{m^2(1+|f|^2)^2}
\]
has local maxima which give the locations $\{(x_a, y_a), \, a=1,..., N\}$ 
of $N$ lumps. The velocities 
$(\dot{x}_a, \dot{y}_a)=(-2m\cos{\theta}/(1+m^2), (1-m^2)/(1+m^2))$ 
are
the same for each lump so (\ref{one_soliton}) should be regarded as a
one--soliton solution. To make contact with the moduli space approximation
write $J_1=\cos{\theta}{\bf 1}+i{\bf a}\cdot {\bf\sm}$ to reveal that
$\cos{\theta}$ measures the deviation of $J_1$ from  the unit sphere in the
Lie algebra ${\bf su}(2)$. If $J$ is initially tangent to the space of
static solutions, then $\cos{\theta}=0$, and we can set
$\mu=i(1+\varepsilon)$, where $\varepsilon\in\R$. The solution is
of the form (\ref{uniton}), but $f$ is rational in 
\[
\om
=z+\varepsilon(z+it)+\frac{\varepsilon^2}{2}\Big(\frac{z-\ov{z}}{2}+it\Big),
\]
so
\[
f(\om)=\frac{(z-Q_1)...(z-Q_N)}{(z-Q_{N+1})...(z-Q_{2N})},
\]
where the $Q$s are linear functions of $(\varepsilon^2\ov{z}, \varepsilon t)$.
The (squared) velocity is
\[
{\cal V}^2=1-4(1+\varepsilon)^2/(1+(1+\varepsilon)^2)^2,
\]
so in the non--relativistic limit (which underlies the moduli space 
approximation) we regard $\varepsilon$ as small. Therefore the 
$Q$s depend only on $t$, and they all move at velocity $\varepsilon$. 
Setting $N=1$, we recover
the charge one solution (\ref{charge1}). More generally we find
that $J$ is given by (\ref{uniton}) with
\[
f=f_2(z)+tf_1(z),
\]
where $f_2=f(\om)|_{\varepsilon=0}$ and 
$f_1=\p f/\p \varepsilon|_{\varepsilon=0}$ are rational functions of 
$z$.

Allowing $\psi$ to have poles of order
higher than one gives solutions which exhibit soliton scattering.
Explicit time--dependent solutions corresponding to scattering 
can be obtained by choosing $\mu_1=i+\varepsilon$, 
$\mu_2=i-\varepsilon$, and taking the limit $\varepsilon\rightarrow 0$.
This yields \cite{W95} 
\be
\label{uniton2}
J_2=\Big({\bf 1}-\frac{2 {p_1}^*\otimes p_1}{||p_1||^2}\Big)
\Big({\bf 1}-\frac{2  {p_2}^*\otimes p_2}{||p_2||^2}\Big),
\ee
where 
\[
p_1=\Big(1, -\frac{i}{2}f_1\Big), \qquad 
p_2=(1+\frac{1}{4}|f_1|^2)\Big(1, -\frac{i}{2}f_1\Big)
-if\Big(\frac{i}{2}\ov{f_1},-1\Big), \qquad f(z, t)=f_2+tf_1',
\]
and $f_1$ and $f_2$ are rational functions of $z$.
In \cite{W95} the 90 degree scattering was illustrated by choosing 
$f_1=2iz, f_2=2iz^2$. More complicated examples were considered
in \cite{I96,DT04}.

It was recently observed \cite{IM04} that the total (kinetic+gradient) energy 
of the solution (\ref{uniton2}) is quantised, and equal to
$8\pi N$, where generically $N=2\deg{f_1}+\deg{f_2}$. However, 
$N=\mbox{max}\;(2\deg{f_1}, \deg{f_2})$ if both $f_1, f_2$ are polynomials.
Therefore for all $t$ the total energy of (\ref{uniton2}) is equal to
the energy (\ref{uniton_en}) of some static solution (\ref{uniton}).

Solutions to (\ref{chiral}) obtained in the moduli space approximation
have energies close to their potential energy (\ref{uniton_en}) as their
kinetic energy is small. We should therefore expect that some of these
approximate solutions arise from (\ref{uniton2}) by a limiting procedure.

To demonstrate how this limiting procedure is achieved first
observe that 
solutions to (\ref{chiral}) are defined up to a multiple by a constant element
of $SU(2)$. The static solution (\ref{uniton}) with $f=f_2$ arises from (\ref{uniton2}) by using this freedom and setting $f_1=0$
\[
\left (
\begin{array}{cc}
i&0\\
0&-i
\end{array}
\right )J_2|_{f_1=0}=
\frac{i}{1+|f_2|^2}
\left (
\begin{array}{cc}
1-|f_2|^2&2f_2\\
2\ov{f_2}&|f_2|^2-1
\end{array}
\right )=J_{\mbox{static}}.
\]
Moreover the energy density  of (\ref{uniton2})
has maxima where $f=f_2+tf_1'=0$. The lumps are located at the zeros 
$z_a=z_a(t)$, $a=1, ..., \deg{f}$ of $f$ and 
the squared velocity of each lump is
\[
{\cal V}_a^2=\frac{|f_1'|^2}{|f_2'+tf_1''|^2}\Biggr|_{z=z_a},
\]
so that $|f_1'|^2$ is small in the non--relativistic  limit. Therefore
$|f_1|$ is also small as we choose $J_2$ to be tangent to the space of 
static solutions at $t=0$. Keeping only the linear terms in $f_1$ 
in (\ref{uniton2}) yields
\[
\left (
\begin{array}{cc}
i&0\\
0&-i
\end{array}
\right )J_2=
\frac{i}{1+|f|^2}
\left (
\begin{array}{cc}
1-|f|^2-i(f_2\ov{f_1}+\ov{f_2}f_1)&2f\\
2\ov{f}&|f|^2-1-i(f_2\ov{f_1}+\ov{f_2}f_1)
\end{array}
\right ).
\]
The term $(f_2\ov{f_1}+\ov{f_2}f_1)$ can also be dropped by rescaling the 
coordinates $x^{\mu}\rightarrow x^{\mu}/\varepsilon$.

Comparing the resulting expression with (\ref{uniton}) will give  
a motion on the moduli space of static solutions
if $f_2, tf_1'$
and $f_1^2$
lie in the common space of rational maps of degree $\deg{f_2}$. 
To achieve this, we therefore take
\[
f_1=\frac{p(z)}{q(z)}, \qquad f_2=\frac{r(z)}{q(z)^2},
\]
where $r$ is of degree $2n$ and $p$ and $q$ are of degree at most $n$.
This is one of the non--generic cases in the analysis of \cite{IM04}, 
and the total energy is equal to $8\pi\deg{f_2}$. The resulting motion
on the moduli space of static solutions of charge $\deg{f_2}$ is given
by (\ref{uniton}) with
\be
f(z, t)=\frac{r+t(p'q-pq')}{q^2}.
\ee
This motion is restricted to a geodesic submanifold as 
the parameters in the denominator of $f$ are fixed.
In particular, setting $q=1$, 
we can take $f_2(z)$ to be a polynomial of degree $2n$
and $f_1(z)$
to be a polynomial of degree at most $n$. 

\section{Conclusions}
\setcounter{equation}{0}
The space of all time--dependent finite energy solutions to (\ref{chiral})
is infinite--dimensional. Restricting to static solutions singles
out finite--dimensional families (\ref{uniton}). In this paper we have
shown that a geodesic motion on the moduli space of static solutions 
approximates the non--relativistic dynamics of Ward solitons.
Some of these approximate solutions have been related to exact uniton
solutions of (\ref{chiral}).

To construct finite--dimensional families of exact time--dependent
solutions to (\ref{chiral}),
the finiteness of energy must be supplemented by other conditions.
Ward \cite{W98} has shown that it is sufficient to assume that
the Higgs field ${\Phi}$ tends to 0, and the solution $\psi$ of the associated
linear problem (\ref{laxpair}) tends to ${\bf 1}$ at spatial infinity of 
each spacelike plane. These conditions hold for all Yang--Mills--Higgs fields
which arise from holomorphic vector bundles over the compactified
twistor space. It would be interesting to understand how these more
general finite energy solutions give rise to a motion on the moduli
space of rational maps.

One expects that the integration of
the equations of motion associated to 
(\ref{reduced_lag}, \ref{metric_one_form}) could perhaps be made 
explicit because of the
integrability of (\ref{Wardeq}).
The conservation of the energy (\ref{energy}), and the
$y$--component of the momentum in field theory will yield two candidates for 
conserved quantities on
the moduli space of static solutions, but they are not sufficient to
ensure the solvability. We have already demonstrated that the kinetic
energy gives rise to a conserved mechanical energy
(\ref{Mmetric}). The analogous procedure applied to the $y$--component
of the momentum with a density
\[
P_y={\mbox{Tr}}(J^{-1}J_yJ^{-1}J_t)
\]
gives, using (\ref{uniton}), 
\[
P=\int{\mbox{Tr}}(J^{-1}J_yJ^{-1}J_t)\d x\d y=\int
\frac{8\;\mbox{Im}{(f_z\dot{\ov{f}})}}{(1+|f|^2)^{2}}\d x\d y=
8\pi\frac{\d}{\d t}\mbox{Re}\;(a+b),
\]
where $a, b$ are given by (\ref{bound_pq}), and
the last equality follows from the application of Green's
theorem along the lines which led to (\ref{OmF0}).
The integrability of (\ref{chiral}) 
guarantees the existence of an infinite sequence of `hidden'
conservation laws not related to the space--time symmetries and the Noether
theorem. It remains to be seen whether these additional symmetries
give rise to conservation laws on the moduli space which 
are sufficient to guarantee integrability 
in the sense of the  Arnold--Liouville theorem.

The WZW term in the Lagrangian generates a magnetic field on the space
of all fields; however, we showed that this vanishes on the moduli
space. The resulting flat connection (\ref{magA}) could still be
interesting, because the moduli space of based rational maps is not
simply connnected. If non--trivial, it would imply that the the 
quantization of Ward
solitons in the low-energy limit differs from the quantization of 
standard sigma model lumps. 

\section*{Acknowledgements}
We thank Michael Atiyah,  Theodora Ioannidou and Maciej Przanowski
for helpful discussions. 

\end{document}